\documentclass[a4paper,portrait,british,aps,groupedaddress,twocolumn,superscriptaddress,citeautoscript,showkeys,nofootinbib, longbibliography]{revtex4-1}

\usepackage{geometry} 
\usepackage{amsmath} 
\usepackage{float} 
\usepackage{xr}
\makeatletter
\newcommand*{\addFileDependency}[1]{
  \typeout{(#1)}
  \@addtofilelist{#1}
  \IfFileExists{#1}{}{\typeout{No file #1.}}
}
\makeatother

\newcommand*{\myexternaldocument}[1]{%
    \externaldocument{#1}%
    \addFileDependency{#1.tex}%
    \addFileDependency{#1.aux}%
}
\myexternaldocument{supp01}
\usepackage{natbib}
\usepackage[table,xcdraw,dvipsnames]{xcolor}
\usepackage{hyperref} 
\usepackage[capitalise, nameinlink]{cleveref} 
\usepackage{matlab-prettifier} 
\usepackage{rotating} 
\usepackage{tabularx} 
\usepackage{dsfont} 
\usepackage[justification=centering]{caption} 
\usepackage[labelsep=space]{subcaption} 
\usepackage{listings}
\usepackage{amssymb} 
\usepackage{wrapfig} 
\usepackage{enumitem} 
\usepackage{graphicx} 
\usepackage[toc,page]{appendix}
\usepackage{multirow} 
\usepackage[nottoc,numbib]{tocbibind} 
\usepackage{subfiles} 
\usepackage{outlines} 
\usepackage{braket} 
\usepackage{esint} 
\usepackage{csquotes} 
\usepackage{epigraph} 
\usepackage{bm} 
\usepackage{blkarray} 
\usepackage{siunitx} 

\usepackage{mathtools} 
\usepackage[bottom]{footmisc} 

\definecolor{lightgrey}{RGB}{200,200,200}

\newcommand{\etal}[1]{\textit{et al.}#1}
 
\newcommand{\silicondioxide}[1]{SiO$_2$#1}
\newcommand{\aluminiumoxide}[1]{Al$_2$O$_3$#1}
\newcommand{\galliumoxide}[1]{Ga$_2$O$_3$#1}

\newcommand{\hafniumdioxide}[1]{HfO$_2$#1}

\newcommand{\mobilityunits}{cm$^2$V$^{-1}$s$^{-1}$}

\newcommand{\fakesection}[1]{
  \par\refstepcounter{section}
  \sectionmark{#1}
  \addcontentsline{toc}{section}{\protect\numberline{\thesection}#1}
}

\bibpunct{[}{]}{,}{s}{}{;}
\allowdisplaybreaks

\hypersetup{
	linkcolor={red!50!black},
	citecolor={blue!50!black},
	urlcolor={blue!80!black},
	pdfborder={0 0 1}
}
\AtBeginDocument{\hypersetup{ 
		pdfborder={10 10 0.1 [1 0]}, 
		citebordercolor={0 0 1}, 
		linkbordercolor={0 0.9 0}, 
}}
\DeclareGraphicsExtensions{.pdf,.png,.jpg,.svg}
\graphicspath{{"./graphics-main2/"}, {"./graphics-supp2/"}}

\newcounter{myfootnotes}



\begin{document}

\title{Mechanically transferred large-area \galliumoxide{} passivates graphene and suppresses interfacial phonon scattering}
\date{\today}
\author{Matthew Gebert}
\affiliation{School of Physics and Astronomy, Monash University, Victoria 3800, Australia}
\affiliation{ARC Centre of Excellence in Future Low-Energy Electronics Technologies, Monash University, Victoria 3800 Australia}

\author{Semonti Bhattacharyya}
\email{Corresponding authors: semonti.bhattacharyya@monash.edu and michael.fuhrer@monash.edu}
\affiliation{School of Physics and Astronomy, Monash University, Victoria 3800, Australia}
\affiliation{ARC Centre of Excellence in Future Low-Energy Electronics Technologies, Monash University, Victoria 3800 Australia}
\altaffiliation{Corresponding author: semonti.bhattacharyya@monash.edu}

\author{Christopher C Bounds}
\affiliation{School of Physics and Astronomy, Monash University, Victoria 3800, Australia}

\author{Nitu Syed}
\affiliation{School of Physics, The University of Melbourne, Melbourne, Parkville VIC 3010, Australia}
\affiliation{School of Engineering, RMIT University, Melbourne, Victoria 3000, Australia}

\author{Torben Daeneke}
\affiliation{School of Engineering, RMIT University, Melbourne, Victoria 3000, Australia}
\affiliation{ARC Centre of Excellence in Future Low-Energy Electronics Technologies, RMIT University, Melbourne, Australia 3000}

\author{Michael S. Fuhrer}
\email{Corresponding authors: semonti.bhattacharyya@monash.edu and michael.fuhrer@monash.edu}
\affiliation{School of Physics and Astronomy, Monash University, Victoria 3800, Australia}
\affiliation{ARC Centre of Excellence in Future Low-Energy Electronics Technologies, Monash University, Victoria 3800 Australia}
\altaffiliation{Corresponding author: michael.fuhrer@monash.edu}

\begin{abstract}
	We demonstrate a large-area passivation layer for graphene by mechanical transfer of ultrathin amorphous \galliumoxide{} synthesized on liquid Ga metal. A comparison of temperature-dependent electrical measurements of millimetre-scale passivated and bare graphene on \silicondioxide{}/Si indicate that the passivated graphene maintains its high field effect mobility desirable for applications. Surprisingly, the temperature-dependent resistivity is reduced in passivated graphene over a range of temperatures below 220 K, due to the interplay of screening of the surface optical phonon  modes of the \silicondioxide{} by high-dielectric-constant \galliumoxide{}, and the relatively high characteristic phonon frequencies of \galliumoxide{}. Raman spectroscopy and electrical measurements indicate that \galliumoxide{} passivation also protects graphene from further processing such as plasma-enhanced atomic layer deposition of \aluminiumoxide{}.\\
	
	Keywords:\hspace{1em}
    ``Chemical vapour deposition (CVD) Graphene",\hspace{1em}
    ``mm-scale oxide dielectric",\hspace{1em}
    ``passivation",\hspace{1em}
    ``remote interfacial polar phonon scattering",\hspace{1em}
    ``van der Waals heterostructure"
    ``dielectric screening"
    ``surface interface interaction"
\end{abstract}
\maketitle

\pagenumbering{arabic}

\fakesection{Introduction} 

Insulating layers are essential components of van der Waals heterostructures~\cite{novoselov_2d_2016} isolating materials electronically, passivating them, and enabling electrostatic gating. High-quality hexagonal boron nitride (h-BN), hand-exfoliated from small single crystals, has been widely used as a wide bandgap insulator for vdW heterostructures, enabling exceptional device quality~\cite{dean_boron_2010, mayorov_micrometer-scale_2011,xue2011scanning, decker2011local}. However, difficulties in producing large-area h-BN of similar quality\cite{lu_experimental_2015, zavabeti_two-dimensional_2020} have so far precluded industrial-scale applications~\cite{novoselov_roadmap_2012, flagship2019european,  C4NR01600A}, prompting a search for other suitable insulators to enable large-area vdW heterostructures.
\begingroup 
\setlength{\unitlength}{1mm}

\definecolor{lgr}{RGB}{167,200,175} 
\definecolor{lbl}{RGB}{128,165,175} 
\definecolor{lpu}{RGB}{184,61,203} 

\definecolor{lgra}{RGB}{90,102,140} 
\definecolor{lsio}{RGB}{190,139,157} 
\definecolor{lsi}{RGB}{173,159,202} 
\definecolor{lau}{RGB}{229,213,112} 

\begin{figure*}
	\centering
    \makebox[\textwidth][c]{
        \begin{picture}(178,76)
        \small
            \put(000,000){\includegraphics{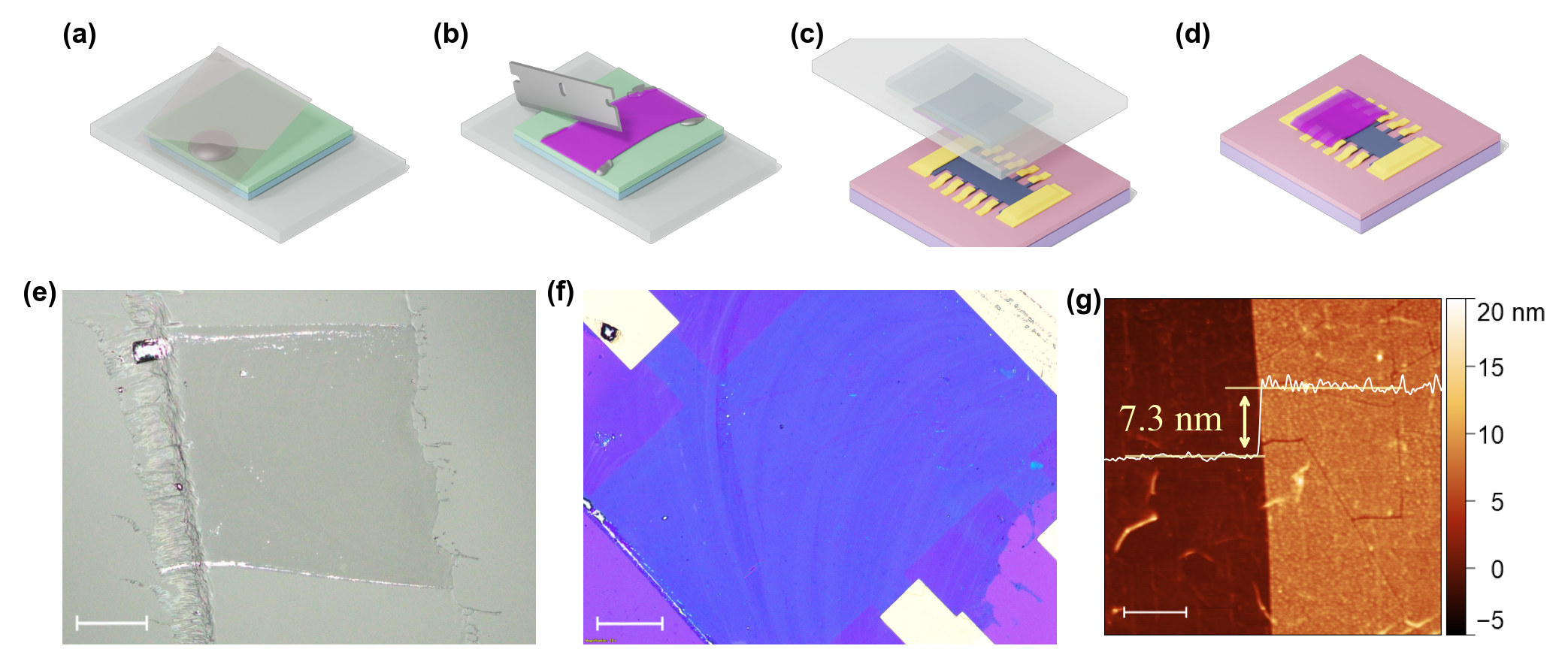}}
    	    \put(000,048){{Gallium droplet}}
    	    \put(012,052){\vector(2,1){10}}
    	    \put(038,048){{Glass slip/slide}}
    	    \put(046,051){\vector(-2,1){16}}
    	    \put(061,049){\vector(4,1){10}}
    	    \put(034,066){{\textcolor{lbl}{GelPak}/}}
    	    \put(039,062){{\textcolor{lgr}{PPC}}}
    	    \put(079,048){{Ga residues}}
	        \put(083,051){\vector(-1,2){5}}
    	    \put(078,064){{\textcolor{lpu}{Ga$_2$O$_3$}}}
	        \put(99,054){{\textcolor{lau}{Au}}}
	        \put(105,049.5){{\textcolor{lgra}{Gr}}}
	        \put(125,056.5){{\textcolor{lsio}{SiO$_2$}}}
	        \put(125,048){{\textcolor{lsi}{Si}}}
	        \put(030,005){{\textcolor{black}{PPC}}}
	        \put(030,018){{\textcolor{black}{\galliumoxide{}}}}
	        \put(047,039){{\textcolor{black}{Gel-Film}}}
	        \put(070,039){{\textcolor{black}{Au}}}
	        \put(066,011){{\textcolor{black}{Gr}}}
	        \put(112,005){{\textcolor{black}{\silicondioxide{}}}}
	        \put(083,005){{\textcolor{black}{\galliumoxide{}}}}
	        \put(138,008){{\textcolor{white}{Gr}}}
	        \put(145,008){{\textcolor{white}{\galliumoxide{}}}}
        \end{picture}
    }
    \caption[Transfer of \galliumoxide{} thin film on Gr.]{Characterization of \galliumoxide{} thin film transferred on Gr. Schematic representation of 
    \textbf{(a) }{Gallium metal positioned to be rolled across a PPC/Gel-Pak polymer stack using a cover slip,}
    \textbf{(b) }{\galliumoxide{} after rolling with some Gallium metal residues, which can be cut away by a razor,}
    \textbf{(c) }{transfer of \galliumoxide{} film onto Gr/\silicondioxide{}/Si device, and}
    \textbf{(d) }{device after removing polymer residues.}
    \textbf{(e)} {Optical darkfield micrograph of \galliumoxide{} on PPC after cutting to size. The silver colored dots are liquid gallium droplets. The \galliumoxide{} sheet at the centre is bordered by liquid gallium. The scale bar is 200 $\mu$m.} 
    \textbf{(f) }{Brightfield optical micrograph of \galliumoxide{} (deep blue sheet on the device) transferred on Gr-device. The scale bar is 50 $\mu$m.} 
    \textbf{(g) }{Topographic image  of \galliumoxide{}-on-Gr sheet obtained by intermittent contact atomic force microscopy (AFM). The left side of the image shows bare Gr, and right side of the image shows \galliumoxide{}-covered Gr. Mean height difference is 7.3 nm as shown in the overlaid line profile, and the scalebar is 2 $\mu$m.}}\label{fig:01-microscopy-afm}
\end{figure*}
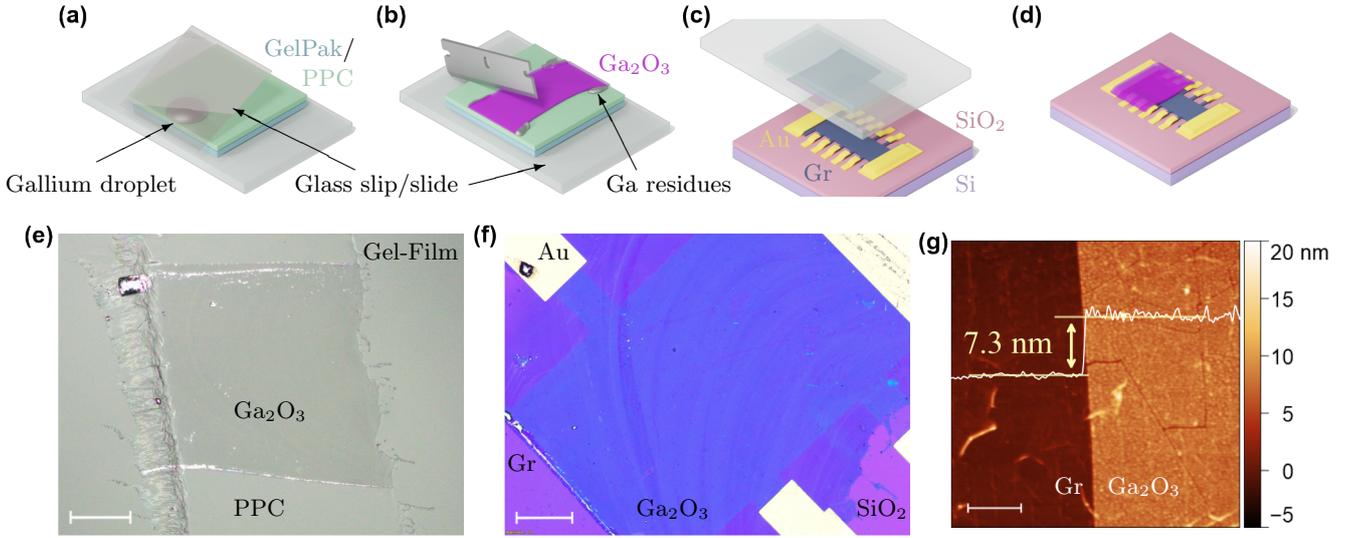
\endgroup

In the case of encapsulating graphene, optimizing the material is highly complex; graphene's electronic properties are largely determined by the insulator's properties, including charged impurity concentration\cite{chen_charged-impurity_2008}, dielectric constant\cite{jang_tuning_2008}, and surface optical (SO) phonons which remotely scatter carriers in the graphene\cite{fratini_substrate-limited_2008,chen_intrinsic_2008}, and trade-offs exist, e.g. insulators with intermediate dielectric constants may be optimal\cite{konar_effect_2010}. 

Recently, the surface of liquid metals has been used to synthesize large-area atomically thin materials with facile mechanical transfer onto other substrates\cite{zavabeti_liquid_2017, kalantar-zadeh_two_2016, daeneke_liquid_2018, D1CS01166A, zavabeti_two-dimensional_2020}. Indeed,  \galliumoxide{} has already been shown to be an excellent encapsulating layer for transition metal dichalcogenide crystals (TMDs) \cite{wurdack_ultrathin_2021}, preserving and even enhancing their optical properties. 

Here we investigate liquid-metal synthesized \galliumoxide{} as a large-area encapsulating layer for graphene with an intermediate relative static dielectric constant $\kappa\sim10$~\cite{passlack_dielectric_1994}. We mechanically transfer large-area (millimeter-scale) \galliumoxide{} onto one portion of a millimeter-scale graphene-on-\silicondioxide{} device, allowing us to compare the electronic transport properties of bare and \galliumoxide{}-encapsulated portions of the same device. We find that coating graphene with \galliumoxide{} preserves the charge carrier mobility close to 3,000 \mobilityunits{}. Surprisingly, we observe a reduction in temperature-dependent resistivity at temperatures below 220 K in the graphene encapsulated by the \galliumoxide{} dielectric, explained by the interplay of screening of the SO phonon modes of the \silicondioxide{} by high-dielectric-constant \galliumoxide{}, and the relatively high characteristic phonon frequencies of \galliumoxide{} itself. We further show that \galliumoxide{} is useful as a passivation layer, protecting against damage from deposition of \aluminiumoxide{} via plasma enhanced ALD. 

\fakesection{Results \& Discussion}


Devices were fabricated (see Section S1 and S2, Supporting Information) using a commercial (Graphene Supermarket) CVD-grown monolayer graphene/monolayer h-BN film (henceforth referred to as ``Gr'') already transferred onto 285 nm \silicondioxide{}/Si  (p-doped) substrate that functions as a global back-gate dielectric and electrode. The Gr was then etched into a Hall bar geometry 0.4 mm wide and 1.2 mm long, with multiple voltage electrodes spaced by 0.25 mm, and contacted by Ti/Au electrodes fabricated using conventional photolithography. Next, mm-scale ultrathin \galliumoxide{} was prepared on a PPC film on a PDMS stamp (Gel-film, Gelpak) through a liquid metal ``squeeze-printing" \cite{D1CS01166A} technique. Finally, \galliumoxide{} was deterministically transferred onto half of the Gr device\cite{wurdack_ultrathin_2021}. We compare the experimental signatures of `bare' and `\galliumoxide{}-covered' parts of graphene in the same Hall bar device to understand the effect of \galliumoxide{}.

\hyperref[fig:01-microscopy-afm]{\Cref*{fig:01-microscopy-afm}} illustrates the steps in the construction of the \galliumoxide{}-on-Gr device. The process of transferring ultrathin \galliumoxide{} films on such Gr-devices are schematically represented in \cref{fig:01-microscopy-afm}a-d). First, a mm-scale ultrathin \galliumoxide{} film was prepared on a PPC film mounted on a PDMS stamp through a liquid metal printing technique known as ``squeeze-printing" \cite{D1CS01166A} (\cref{fig:01-microscopy-afm}a). This film was then cut into approprate size to cover half of the Gr-device as well as to get rid of additional Ga-particles (\cref{fig:01-microscopy-afm}b), and was finally deterministically transferred onto half of the Gr Hall bar device using a home-made van der Waals stacking set up \cite{wurdack_ultrathin_2021} (\cref{fig:01-microscopy-afm}c and d).

\hyperref[fig:01-microscopy-afm]{\Cref*{fig:01-microscopy-afm}e}) shows a dark-field image of a squeeze-printed \galliumoxide{} film on PPC/PDMS assembly. This film was trimmed to 0.7 mm $\times$ 0.65 mm to match the Gr-device. The darkfield image highlights the Ga-metal residue, left from the squeeze-printing processs, which is negligible in the interior area of the film, and mostly appears at the boundary.
 \hyperref[fig:01-microscopy-afm]{\Cref*{fig:01-microscopy-afm}f}) shows a bright-field optical image of the \galliumoxide{} film transferred on the Gr-device. The optical contrast of the amorphous \galliumoxide{} film is largely uniform, though slight variations are visible, indicating similar thickness across the thin film (Section S9, Supporting Information). 
\hyperref[fig:01-microscopy-afm]{\Cref*{fig:01-microscopy-afm}g)} shows an atomic force micrograph of both \galliumoxide{}-covered (right-half) and bare side (left-half) of a Gr-device. The AFM line profile (overlaid) yields a step height of 7.3 nm for \galliumoxide{}, similar to AFM measurements performed on similar devices (Section S9, Supporting Information), 
and consistent with thicknesses reported by Wurdack \etal{}\cite{wurdack_ultrathin_2021}.

\begin{figure*}
	\centering
    \makebox[\textwidth][c]{\includegraphics{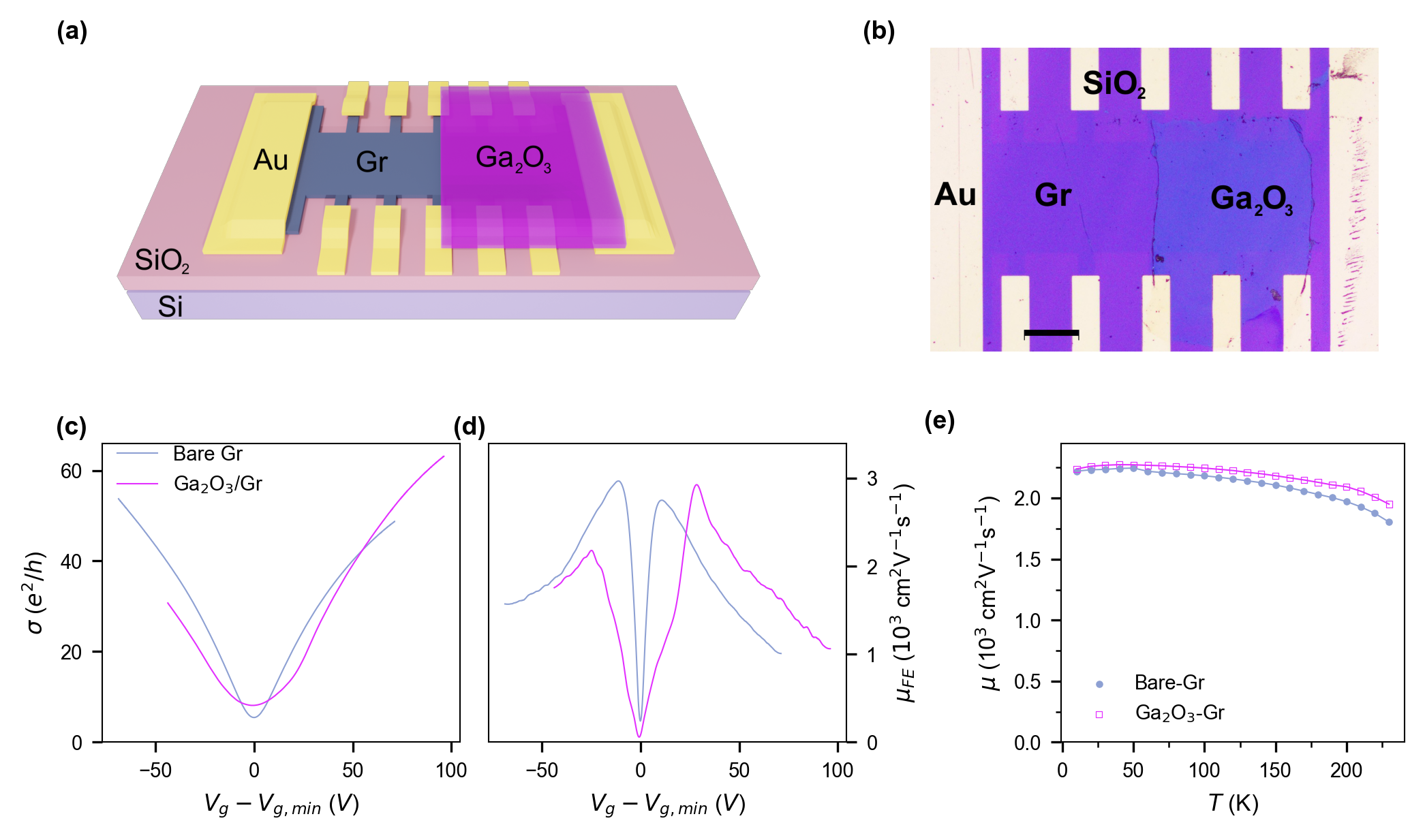}} 
    \caption[Gate voltage-dependent electrical transport measurements of \galliumoxide{}-covered and bare Gr field effect devices.]
    {Gate voltage and temperature dependent electrical transport measurements of \galliumoxide{}-covered and bare Gr field effect devices. 
    \textbf{(a)} Schematic illustration and \textbf{(b)} optical microscope image of a Gr-device after \galliumoxide{} transfer. The scale bar is 200 $\mu$m. 
    \textbf{(c)} Longitudinal conductivity $\sigma$ and  \textbf{(d)} field-effect mobility $\mu_{\text{FE}}$ as a function of gate voltage $V_g$ offset by gate voltage at minimum conductivity $V_{g,min}$, for both bare Gr and \galliumoxide{}-covered Gr at 100 K. 
    \textbf{(e) }{Temperature ($T$)-dependent mobility ($\mu$) calculated at charge carrier density $n=5 \times 10^{12}$ cm$^{-2}$.}}\label{fig:02-transport-device}
\end{figure*}


\Cref{fig:02-transport-device} compares the gate-voltage and temperature dependent electrical transport properties of the bare and \galliumoxide{}-covered Gr-devices.
The 3D schematic of the device and top-view micrograph are shown in \hyperref[fig:02-transport-device]{\Cref*{fig:02-transport-device}a) and b}) respectively. Figure 2b) shows that the transferred \galliumoxide{} film covers half of the Gr-device. The orientation of the \galliumoxide{} film has been carefully controlled so that the Ga-metal particles at the boundary do not affect the electrical transport characteristics between the voltage probes.

\hyperref[fig:02-transport-device]{\Cref*{fig:02-transport-device}c)} shows the gate voltage ($V_g$)-dependence of the longitudinal conductivity $\sigma$ measured at temperature \textit{T} = 100 K for both bare and \galliumoxide{} covered Gr. The gate voltage $V_g$ is offset by the gate voltage of minimum conductivity ($V_{g,min} =$ -1.2 V and -26.2 V respectively for the bare and \galliumoxide{}-covered sides) to faciliate comparison between the two sides of the device. The comparison of the two parts of the sample shows three notable dfferences, with the \galliumoxide{}-covered part showing (i) slightly enhanced conductivity at high $V_g$,  (ii) increased magnitude of minimum conductivity $\sigma_{min}$, and (iii) a broader minimum-conductivity plateau.

 In order to highlight the first of these features we plot the field effect mobility ($\mu_{\text{FE}}=\frac{1}{c_g}\frac{\partial \sigma}{\partial V_g}$; ${c_g}=$ Capacitance of the \silicondioxide{} back gate) obtained from $\sigma(V_g)$ data (\hyperref[fig:02-transport-device]{\Cref*{fig:02-transport-device}d}). The peak electron mobility is slightly improved in the \galliumoxide{}-covered graphene ($\mu=2900$ \mobilityunits{}) relative to bare graphene ($\mu=2800 $ \mobilityunits{}). The highest observed hole mobility in \galliumoxide{}-covered Gr ($\mu=2200$ \mobilityunits{}) is lower than in bare Gr ($\mu=3000$ \mobilityunits{}) but might not be a global maximum, as it is at the edge of the $V_g$ measurement window. $\mu_{\text{FE}}$ is observed to be mostly unchanged by the addition of \galliumoxide{}, if not a little increased at high positive $V_g$. 
 
 The slight enhancement of mobility after deposition of \galliumoxide{} is remarkable because previous experiments found that deposition of oxide usually degrades the mobility of graphene~\cite{liao_graphene-dielectric_2010} due to introduction of disorder. In contrast, screening by a clean dielectric can reduce charged impurity scattering\cite{jang_tuning_2008, newaz_probing_2012}. To infer the impurity concentration in \silicondioxide{}/Gr/\galliumoxide{} with RPA-Boltzmann theory\cite{adam_self-consistent_2007} we use $\kappa=10$ for the amorphous \galliumoxide{}\cite{passlack_dielectric_1994}, and $\mu=2900$ cm$^{2}$V$^{-1}$s$^{-1}$ and calculate an impurity concentration of $n_{imp,\text{Ga}_2\text{O}_3} = 3.8 \times 10^{12}$ cm$^{-2}$ (Section S6.A in the Supporting Information), roughly twice the impurity concentration inferred for our bare graphene on \silicondioxide{} ($n_{imp,bare}=1.8 \times 10^{12}$ cm$^{-2}$). This indicates that our liquid-metal synthesized and mechanically transferred \galliumoxide{} layer has low charged impurity concentration, comparable to thermally grown \silicondioxide{}. 
 
 According to the RPA-Boltzmann theory, the increased $n_{imp}$ should also lead to a very weak reduction of $\sigma_{min}$ \cite{adam_self-consistent_2007} and a narrowing of the minimum conductivity plateau in the \galliumoxide{}-covered graphene (Section S6.B in the Supporting Information), in contrast to our observation (\Cref{fig:02-transport-device}c). The increased $\sigma_{min}$, as well as broadening of the minimum conductivity plateau likely instead reflect additional macroscopic inhomogeneity of the sample~\cite{blake2009influence} induced by the \galliumoxide{}.





In order to further explore the modification of electrical transport in \galliumoxide{}-covered graphene we plotted $\mu$ = \( \frac{\sigma}{ne} \) calculated at $n=5\times10^{12}$ cm$^{-2}$ for both bare and  \galliumoxide{}-covered graphene (\Cref{fig:02-transport-device}e). We observe a gradual reduction in mobility with increasing temperature between 60 K to $\approx 220$~K in both. The overall decline of mobility indicates a temperature-dependent resistivity contribution, which surprisingly appears larger in bare compared to \galliumoxide{}-covered graphene, in contrast to previous experiments where addition of an oxide layer on graphene increased the temperature-dependent resistivity\cite{zou_deposition_2010}.


We expect that dielectric layers affect the temperature dependent mobility of graphene through scattering of charge carriers by SO phonons. This process, also known as remote optical phonon (ROP) scattering, is expected to contribute a resistivity proportional to a Bose-Einstein  distribution \cite{chen_intrinsic_2008}. 
\begin{align}\label{eqn:ROP}
    \rho_{ROP}(V_g,T) = V_g^{-\alpha}\sum_{i}^{M}\dfrac{\beta_i}{e^{\hbar\omega_i/k_B T}-1}
\end{align}
Here the $i^{\text{th}}$ SO mode is described by the SO phonon energy $\hbar\omega_i$ (meV) and respective coupling strength $\beta_i$ ($V_g^{\alpha}$). Empirically the dependence on gate voltage is found to follow a power law with $\alpha\approx$ 1.

To examine the differences in ROP scattering for bare and \galliumoxide{}-covered graphene, we extracted $\rho_{ROP}(V_g,T)$ from $\rho(V_g,T)$ (see Section S10, Supporting Infromation for details). Briefly, we perform a global fit of $\rho(V_g,T)$ at different $V_g$ at 70 K $\leq T \leq 100$ K to determine the acoustic phonon scattering contribution $\rho_{LA}(T)$ which is linear in temperature and independent of $V_g$, and the impurity contribution $\rho_{imp}(V_g, T=0$ K) which depends on $V_g$ but not temperature. Subtracting these two quantities from $\rho(V_g,T)$ allows us to extract $\rho_{ROP}(V_g,T)$.

\begin{figure*}
	\centering
    \makebox[\linewidth][c]{
	\includegraphics[scale=0.65]{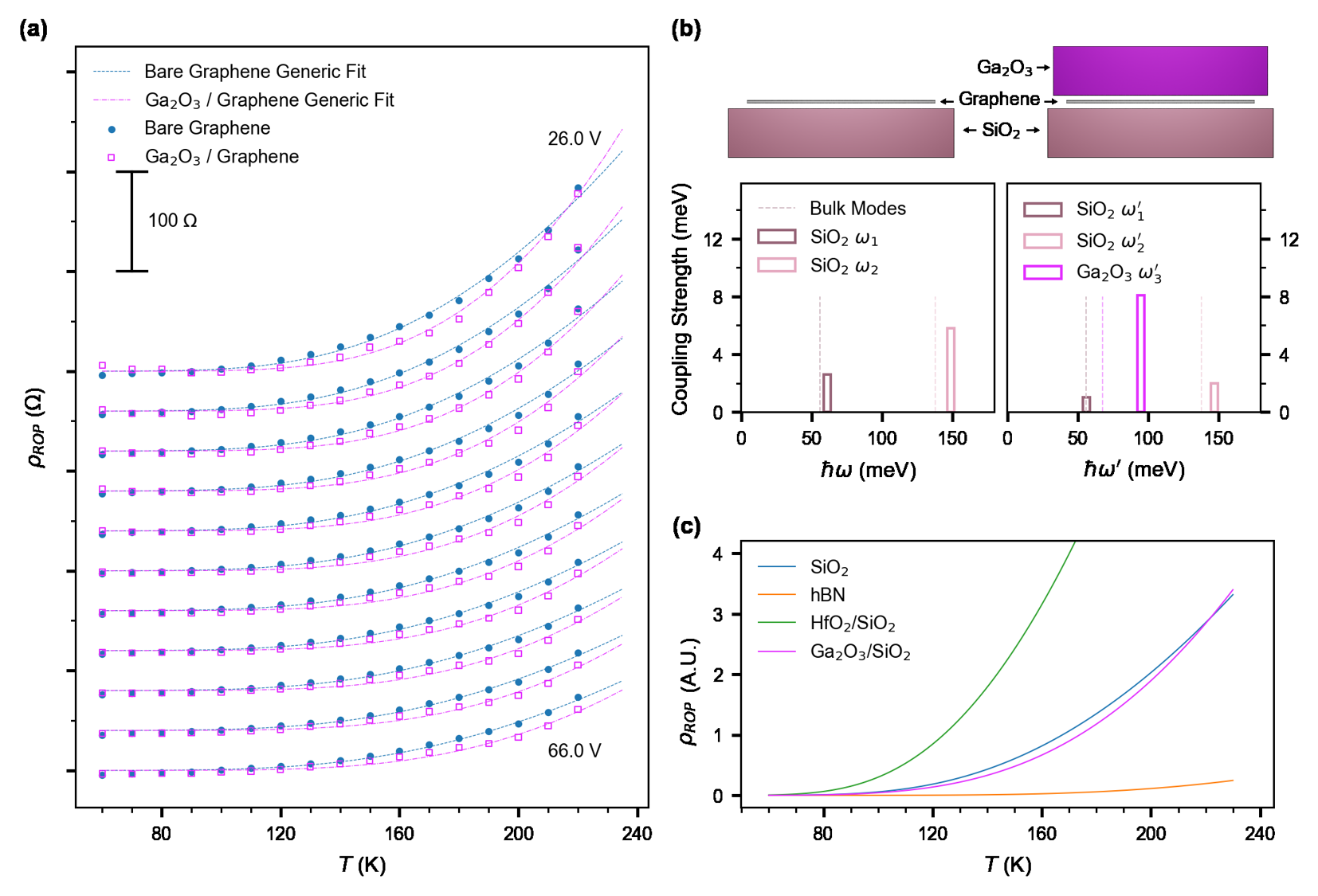}}
	\caption[\textbf{ROP scattering}]{\textbf{Remote optical phonon (ROP) Scattering} \textbf{(a)} ROP contributions to resistivity ($\rho_{ROP}$) extracted from the temperature ($T$)-dependence of resistivity ($\rho$) in bare (filled) and \galliumoxide{}-covered (hollow) graphene. Fits using \cref{eqn:ROP} are plotted with dashed line for bare and dot-dashed line \galliumoxide{}-covered Gr. Each gate voltage, from 26 V through 66 V via 4 V steps, $\rho_{ROP}$ is offset by 40 $\Omega$. \textbf{(b)} Computed frequency and coupling strength of SO phonons in bare and \galliumoxide{}-covered Gr. Dashed lines indicate the corresponding bulk mode phonon frequency. \textbf{(c)} Modelled ROP scattering contribution to the resistivity ($\rho_{ROP}$) in graphene for different dielectric systems. }\label{fig:04-phonons}
\end{figure*}
\Cref{fig:04-phonons}a) shows the contribution of ROP scattering $\rho_{ROP}(V_g,T)$ to resistivity as a function of temperature, for various positive gate voltages (offset from the minimum $V_{g,min}$). $\rho_{ROP}(V_g,T)$ increases super-linearly in temperature, with larger magnitude at smaller $V_g-V_{g,min}$. Remarkably, in the temperature range 70 - 220~K, $\rho_{ROP}$ is lower in the \galliumoxide{}-covered Gr compared to bare Gr for all values of $V_g-V_{g,min}$.
The dashed lines in \Cref{fig:04-phonons}a) are global fits of the data to \cref{eqn:ROP} for a single phonon mode ($M=$1) for bare (dashed) and \galliumoxide{}-covered (dot-dash) graphene.

The results of the fits in \Cref{fig:04-phonons}a) are summarized in Table I. We determine the SO mode energy, $\hbar\omega_0$, to be larger in \galliumoxide{}-covered graphene ($92.8$ meV) compared to bare graphene on \silicondioxide{} ($69.5$ meV). The observation of slightly higher $\hbar\omega_0$ compared to the expected lowest phonon mode for \silicondioxide{}  (61 meV) was also observed by Chen et al.\cite{chen_intrinsic_2008}, and likely due to additional contribution of the higher-energy \silicondioxide{} mode. 
The power-law exponent $\alpha$ is similar for bare and \galliumoxide{}-covered graphene, and close to that of previous literature \cite{chen_intrinsic_2008, zou_deposition_2010}. The coupling strength
$\beta$ is found to be 6.0 ($h/e^2$) for bare graphene, which is roughly double the value of 3.26 ($h/e^2$) found by Chen \etal{}\cite{chen_intrinsic_2008} Due to the different values of $\alpha$ for \galliumoxide{}-covered graphene, it is difficult to directly compare the coupling $\beta$ which has different dimensions and consequently has a different magnitude to that for bare graphene. 




Our observations (Table I) indicate that the smaller  magnitude of $\rho_{ROP}(V_g,T)$ at low temperatures ($T\lesssim220$ K) for \galliumoxide{}-covered graphene is due to a higher effective phonon energy, resulting in a lower $\rho_{ROP}(V_g,T)$ at low temperatures due to lower phonon population, and eventually crossing over to higher $\rho_{ROP}(V_g,T)$ at high $T$ due to stronger coupling. To better understand the lower $\rho_{ROP}$ contribution in \galliumoxide{}-covered graphene, we develop a simple analytical model of the \silicondioxide{}/graphene/\galliumoxide{} heterostructure, following the methodology used in semiconductor inversion layers \cite{fischetti_effective_2001} and graphene systems \cite{fratini_substrate-limited_2008,zou_deposition_2010}. The details of the model are described in the Section S8 Supporting Information.

The resultant phonon frequencies and coupling constants of both bare and \galliumoxide{}-covered graphene is schematically represented in \cref{fig:04-phonons}b) (see Table S2, S3-S6, Supporting Information for more details). We find that the graphene/\silicondioxide{} structure has two SO modes i.e., $\hbar\omega_1$ = 61 meV and $\hbar\omega_2$ = 149 meV. 
\galliumoxide{}/graphene/\silicondioxide{} structure has three SO modes, with energies and coupling strengths shown in \cref{fig:04-phonons}b). Here $\hslash\omega_1' = 56$ meV, $\hslash\omega_2' = 147$ meV are the perturbed \silicondioxide{} modes, while $\hslash\omega_3' = 95$ meV originates in \galliumoxide{}. While all these modes are thermally activated, the \galliumoxide{} mode couples particularly strongly, also reflected in the high disparity between $\epsilon_{Ga_2O_3}^{0}$ and $\epsilon_{Ga_2O_3}^{\infty}$ (Table S2 and Section S8, Supporting Information). At the same time however, the large $\epsilon_{Ga_2O_3}^{\infty}$ screens the ROP scattering from \silicondioxide{} modes. Hence we expect the $\omega_3'$ mode to dominate the temperature-dependent resistivity. Importantly, the energy corresponding to the $\omega_3'$ mode matches well with the $\hbar\omega_0$ value obtained from fitting our experimental data (\cref{fig:04-phonons}a) and Table I.
\begin{table*}[ht]
	\centering
	\caption{Parameters for fits of data in \Cref{fig:04-phonons} for bare and \galliumoxide{}-covered graphene to \Cref{eqn:ROP}}\label{tbl:phonon-fit}
    \begin{ruledtabular}
    		\begin{tabular}{lccc}
    			&               $\alpha$                        	& $\beta$ (V$^{-\alpha}$ $h/e^2$) & $\hbar\omega_0$ (meV) \\\hline
    			Bare			& $\num{0.97} \pm \num{0.03}$	& $\num{6.0}\pm \num{1.1}$ 	& $\num{69.5}\pm \num{1.3}$ \\\hline
    			\galliumoxide{}	& $\num{1.18} \pm \num{0.03}$    & $\num{43.1}\pm \num{6.7}$ 	& $\num{92.8}\pm \num{2.6}$	\\
    		\end{tabular}%
    \end{ruledtabular}
\end{table*}

\Cref{fig:04-phonons}c) shows the analytically obtained $\rho_{ROP}(T)$ for the \galliumoxide{}/graphene/\silicondioxide{} structure and bare graphene/\silicondioxide{} calculated using \cref{eqn:ROP}, using the SO modes as shown in \cref{fig:04-phonons}c) (also given in Table S3 and Table S6, Supporting Information). Also shown for comparison are graphene/h-BN, and HfO$_2$/graphene/SiO$_2$ (Table S4, S5, Supporting Information). We see that $\rho_{ROP}(T)$ in \galliumoxide{}/graphene/\silicondioxide{} is lower than for bare graphene/\silicondioxide{} at temperatures below approximately 220 K. Because ROP is thermally activated, at low temperatures the \silicondioxide{} mode ($\omega_1$) will dominate $\rho_{ROP}$ for bare graphene/\silicondioxide{}, while for the \galliumoxide{}/graphene/\silicondioxide{} structure, the \galliumoxide{} effectively screens these \silicondioxide{} contributions.
At higher $T$, the higher mode energy ($\omega_3'>\omega_1'$) \galliumoxide{} ROP scattering becomes active and quickly begins to dominate due to the higher coupling.
Our observation that addition of \galliumoxide{} to graphene/\silicondioxide{} can lower the overall interfacial phonon scattering may inspire the design of other heterostructures to further reduce scattering phenomena of charge carriers in graphene, perhaps yielding a more significant improvement at or above room temperature.

\begin{figure*}[t]
	\centering
	\makebox[\textwidth][c]{\includegraphics{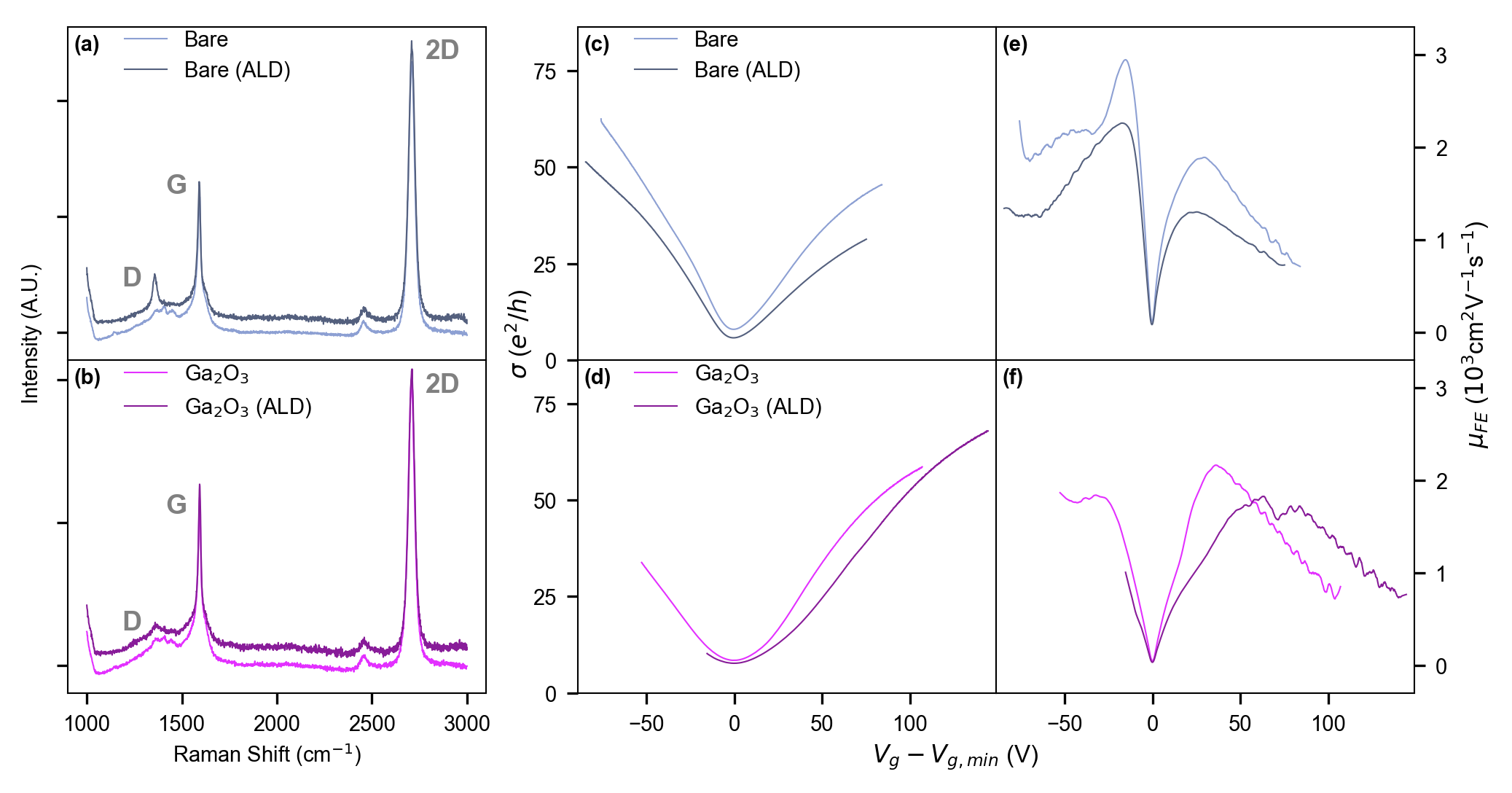}}
	\caption{\galliumoxide{} as a protective layer on graphene against plasma-enhanced atomic layer deposition (ALD) of \aluminiumoxide{}. Raman spectroscopy with D, G and 2D peaks indicated for \textbf{(a)} bare graphene and \textbf{(b)} \galliumoxide{}-covered graphene. Gate-dependence of conductivity ($\sigma$) and mobility ($\mu_{FE}$) respectively for \textbf{(c, e)} bare Gr and \textbf{(d, f)} \galliumoxide{}-covered Gr. Data is shown for the same samples before (lighter shade) and after (darker shaden) of the ALD process. Raman and transport data are taken from different samples at room temperature.}\label{fig:06-ald}
\end{figure*}

In our model, we have ignored the effect of the monolayer h-BN in between graphene and \silicondioxide{}. The model agrees well with experiment and past literature~\cite{chen_charged-impurity_2008,zou_deposition_2010}, suggesting that the surface modes of h-BN/\silicondioxide{} are comparable to that of bare \silicondioxide{}. This may be explained by the similar dielectric constants of the \silicondioxide{} and h-BN resulting in similar SO properties for the monolayer h-BN/\silicondioxide{} composite to bare \silicondioxide{}. Our model on \silicondioxide{}/graphene/\hafniumdioxide{} device matches with previous results as expected~\cite{zou_deposition_2010}. 

Having demonstrated that \galliumoxide{} does not enhance impurity scattering in graphene and even reduces the impact of phonon scattering in a certain temperature range, we now investigate whether \galliumoxide{} is effective in protecting graphene from further processing. Methods of growing large-area dielectric films, such as CVD, atomic layer deposition (ALD), sputtering, e-beam evaporation have proven to be damaging for graphene, leading to enhancement of impurity scattering, and consequently degradation of mobility \cite{fallahazad_dielectric_2010, zou_deposition_2010, tang, jin_study_2009, dlubak_are_2010, maneshian_influence_2011}. 

Our gentle transfer technique for thin \galliumoxide{} avoids such damaging processes, as demonstrated in \Cref{fig:02-transport-device}. We further demonstrate the protective nature of \galliumoxide{} by using plasma-enhanced ALD to grow a 5.5 nm layer of \aluminiumoxide{} over the entire sample, following the method of Tang \etal{} \cite{tang} (see Section S3, Supporting Information for further details), in order to replicate conditions that normally might damage graphene. The ALD chamber temperature is modified to 150$^\circ$ C, to avoid possible change of morphology of amorphous \galliumoxide{} by thermal annealing~\cite{wurdack_ultrathin_2021}.

\Cref{fig:06-ald} compares the effect of \aluminiumoxide{} deposition on the bare and the \galliumoxide{}-covered side of the Gr device. \hyperref[fig:06-ald]{\Cref{fig:06-ald}a) and b)} show Raman spectra on bare and \galliumoxide{}-covered sides respectively before and after ALD deposition. Both areas of the device show monolayer graphene G \& 2D Raman peaks at 1591 cm$^{-1}$ and 2709 cm$^{-1}$, with expected ratio $\sim$2. Before ALD deposition the graphene spectra on both sides are nearly identical, confirming that \galliumoxide{}-transfer process does not lead to any structural disorder in graphene. However, there is a stark difference between the Raman spectra on both sides of the devices after ALD processing, where the bare graphene area shows a fully formed D-peak (1357 cm$^{-1}$), which is nearly absent in pre-processed graphene, and remains unchanged in the \galliumoxide{}-covered graphene even after processing.
The D peak is activated by point disorder, and indicates deposition-induced damage in bare graphene which is not present in \galliumoxide{}-covered graphene. This clearly indicates that \galliumoxide{} transfer has no adverse effect on graphene, and also acts as an encapsulating layer to protect graphene against further damage during subsequent deposition processes. 

This is further supported through electrical transport data obtained on both sides of an identically prepared device before and after \aluminiumoxide{} deposition (\cref{fig:06-ald}c-f).
{\Cref{fig:06-ald}c,d)} and {\Cref{fig:06-ald}e,f)} show the relative change in $\sigma$ and $\mu_{FE}$ respectively before and after ALD, both on bare and \galliumoxide{}-covered graphene. 
After ALD processing, there is a global decrease in $\sigma$ and $\mu_{FE}$ for bare graphene, and peak $\mu$ drops by $\approx30\%$, compared to $<15\%$ in \galliumoxide{}-covered graphene. At higher gate voltages ($V_g - V_{g,min} > 50$ V) the transport in processed \galliumoxide{}-covered graphene shows similar performance to prior to ALD, with higher $\mu_{FE}$ values, and similar $\sigma$ at $\approx100$ V. 
The broadening (and slight reduction) of the $\mu_{FE}$ peak in the \galliumoxide{}-covered side (Fig. \ref{fig:06-ald}f) appears to have been caused by further enhancement of the inhomogeneity already present in the system, as the higher mobility at high carrier density indicates that ALD on \galliumoxide{}-covered graphene does not induce additional impurities, in agreement with the Raman spectroscopy results.

Our results demonstrate that liquid-metal sythesized \galliumoxide{} is a viable large-area mechanically transferred passivation layer for vdW heterostructures. Encapsulation of graphene by \galliumoxide{} preserves the mobility, and reduces ROP scattering in graphene below $T$ = 220 K due to the interplay of high energy phonon modes and dielectric screening in this oxide with intermediate dielectric constant. The large area passivation capability of \galliumoxide{} enables other deposition methods without causing damage at the interface, which should allow integration with a variety of materials and processes. The liquid metal printing technique is highly versatile with a wide range of materials already demonstrated~\cite{D1CS01166A}, hence this work opens the possibility of expanding to other liquid metal printed ultrathin materials for large-area vdW heterostructures.

\section*{Acknowledgements}
This work was supported by the ARC Centre of Excellence in Future Low-Energy Electronics Technologies (CE170100039). We acknowledge Dr. Kaijian Xing for discussions regarding dielectric constant of \galliumoxide{} and Mr. Matthias Wurdack for discussions regarding the transfer process of \galliumoxide{}.

This work was performed in part at the Melbourne Centre for Nanofabrication (MCN) in the Victorian Node of the Australian National Fabrication Facility (ANFF).

\bibliography{./bibtex/MyLibrary2}
\end{document}